\documentclass[final,5p,times,twocolumn]{elsarticle}

 \usepackage{graphics}
 \usepackage{epsfig}

\usepackage{amssymb}

\begin{document}

\title {\bf{Enzo Flaminio and Neutrinos}}

\author{Giorgio Giacomelli\\ 
Physics Department of the University of Bologna and INFN Sez. of Bologna~~{\footnotesize (giacomelli@bo.infn.it)}\\
\vspace{0.5cm}
{\bf \large{Enzo's Festa, Pisa 11 december 2009.}}}

\begin{frontmatter}

\begin{abstract}
The research activities in collaboration with Enzo Flaminio are summarized and discussed. The first experiment was the monochromatic $K^{0}_{L}$ interactions in the 2m CERN hydrogen bubble chamber. The second was $\nu_{\mu}$ and $\overline{\nu_{\mu}}$ in the BEBC bubble chamber filled with deuterium: the main emphasis was on the determination of the $\nu_{\mu}$N and $\overline{\nu_{\mu}}$N structure functions and on their dependence on Q$^{2}$ and x. We cooperated in the initial period of the MACRO experiment at Gran Sasso, in particular in the searches for magnetic monopoles, neutrinos from gravitational stellar collapses and on oscillations of atmospheric neutrinos. Now we are cooperating in the Antares neutrino telescope.  
\end{abstract}

\end{frontmatter}

\section{Introduction.} 
\label{intro}
I met Enzo Flaminio long time ago in the US. We both performed experiments with bubble chambers and with electronics detectors and collaborated in several experiments using the two techniques.

We started collaborating in an experiment with the 2m CERN hydrogen bubble chamber, and made data compilations of the results, and we continued the collaboration studying $\nu_{\mu}$ and $\overline{\nu_{\mu}}$ broad band beam interactions in the BEBC bubble chamber filled with deuterium. 

When the bubble chamber era was ending we proposed, built and used the MACRO detector at the Gran Sasso underground laboratory.

Now we are collaborating in the ANTARES neutrino telescope in the Mediterranean sea \cite{1}.

\section{Bubble Chambers.}

The bubble chambers made their major contributions to particle physics from the 1950s until the 1980s \cite{2,3}. In Italy the bubble chamber technique lead to a revival of fundamental research in particle physics, with profitable cooperations between Departments of Physics and Sections of INFN (the National Institute for Nuclear and Subnuclear Physics). The CNAF-INFN center in Bologna played a central coordinating role for precise measurements and for central computing.

Large bubble chambers were built and run by experts in large laboratories like CERN and used refined beams at accelerators of increasing energy. These chambers were considered facilities that could be used by many groups, and this increased the number of international collaborations, with groups from different countries, about 20-50 physicists per experiment. The role of large laboratories like CERN was always a central one and the large CERN bubble chambers took tens of thousands of stereoscopic pictures. Computer technology grew in parallel with the increase in size and automation of the bubble chambers. The installation of mainframe computing capacity was driven by the demands of bubble-chamber physics.

\begin{figure}[h]
\begin{center}
{\centering\resizebox*{!}{10cm}{\includegraphics{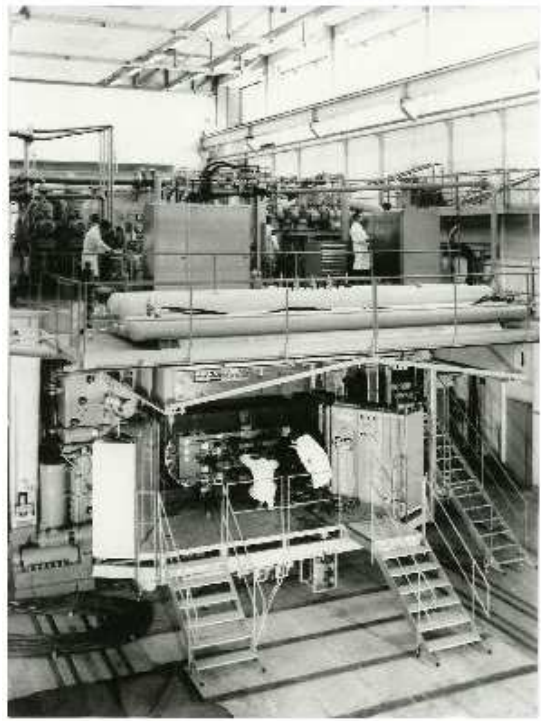}}\par}
\caption{\small The 2m CERN hydrogen bubble chamber.}
\label{fig:1}
\end{center}
\end{figure}

\begin{figure}[h]
\begin{center}
\vspace{-0.3cm}
\hspace{-0.5cm}
{\centering\resizebox*{!}{6cm}{\includegraphics{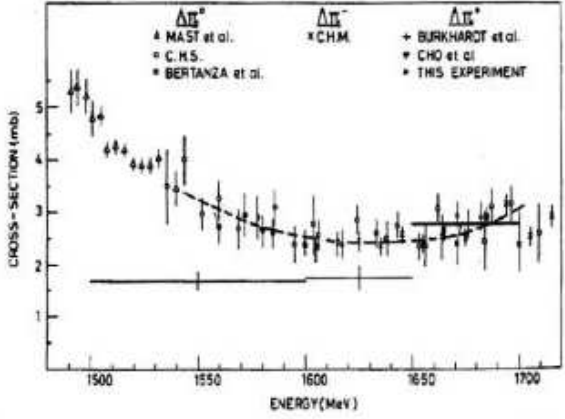}}\par}
\caption{\small $K^{O}_{L}$ experiment: $\Lambda$$\pi$ cross sections. The dashed line is a fit to all the data.}
\label{fig:2}
\end{center}
\end{figure}

\begin{figure}[h]
\begin{center}
{\centering\resizebox*{!}{6.5cm}{\includegraphics{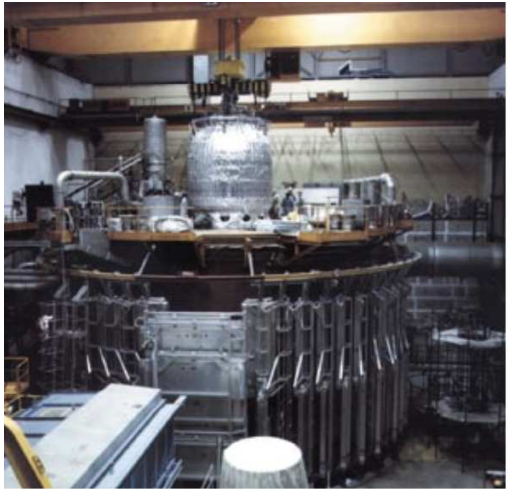}}\par}
\caption{\small The 3.7m CERN BEBC. Notice the ``body'' of the chamber being inserted and the ``picket fence'' of the external muon identifier. BEBC was filled with H$_{2}$, D$_{2}$ or heavy liquids.}
\label{fig:3}
\end{center}
\end{figure}

The first experiment in collaboration with Flaminio was a bubble chamber experiment with a monochromatic/dichromatic $K^{0}_{L}$ beam in the 2m CERN bubble chamber filled with liquid hydrogen, Fig. 1 \cite{4,5,6}. The main purpose was to perform experiments with $K^{0}_{L}$ with precisions and statistical accuracies typical of bubble chamber experiments using charged hadron beams. Data were taken at 13 different incident $K^{0}_{L}$ lab. momenta in the range 300-800 MeV/c. The bubble chamber film was scanned for a visible $V^{0}$. The data concerned the strong interaction channels :

\begin{equation}
K^{0}_{L}p \rightarrow \Lambda^{0}\pi^{+}, ~~\rightarrow \Sigma^{0} \pi^{+}, ~~\rightarrow \Lambda^{0}\pi^{+}\pi^{0}, ~~\rightarrow K^{0}_{s}p  
\end{equation}
Fig. 2 shows a compilation of cross sections for $\Lambda\pi$ production, including $K^{0}_{L}p \rightarrow \Lambda^{0}\pi^{+}$. The goals of the experiment were reached and the data result were added to data compilations \cite{6}.

\vspace{0.5cm}
The second experiment was a major bubble chamber experiment with the $\nu_{\mu}$ and $\overline\nu_{\mu}$ broad band beams from the CERN SPS in the BEBC 3.7 m bubble chamber filled with liquid deuterium, see Fig. 3. More than 30000 interactions were measured [7-27].

\begin{figure}[h]
\begin{center}
\vspace{-0.3cm}
\hspace{-0.5cm}
{\centering\resizebox*{!}{10cm}{\includegraphics{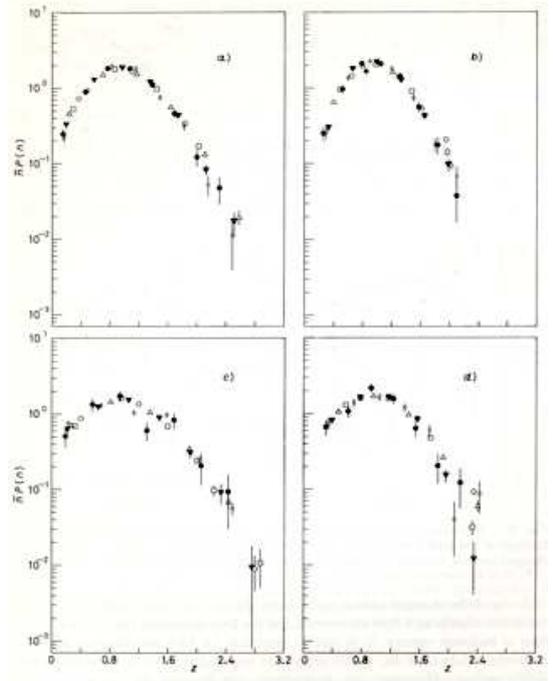}}\par}
\caption{\small $\nu_{\mu}d$, $\overline{\nu_{\mu}}d$ interactions in BEBC. Charged hadron multiplicity distributions in the KNO form for different hadronic energies.}
\vspace{-0.5cm}
\label{fig:4}
\end{center}
\end{figure}

 The first results concerned the multiplicity distributions of charged hadrons produced in $\nu_{\mu}$p, $\nu_{\mu}$n, $\overline{\nu_{\mu}}$p, $\overline{\nu_{\mu}}$n Charge Current (and also Neutral Current) interactions in the E$_{\nu}$ lab. energy range 5-150 GeV, corresponding to the produced hadronic energy range 2-14 GeV. The experimental distributions were analyzed in terms of the binomial distribution in the KNO form, showing also the validity of KNO scaling, Fig. 4. (Fig. 4, Fig. 5 and Fig. 6 are only meant to give examples of the topics studied and of the statistical precisions obtained.). The measured CC interactions were used to obtain the complete set of structure functions on protons and neutrons and the x and Q$^{2}$ dependences of the structure functions of up and down valence quarks and antiquarks; finally a QCD analysis yielded the four non singlet structure functions $xF_{3}^{\nu N}$, $xu_{v}$, $xd_{v}$ and $F_{2}^{\nu n}$ $F_{2}^{\nu p}$, see Fig. 5. 

The NC chiral coupling constants were also measured in neutrino and antineutrino interactions on protons and neutrons, see Fig. 6 (Assuming $\rho$=1 we obtained the corrisponding value of $sin^{2}\theta_{W}$=0.247$\pm$0.018$\pm$0.023). 

Among the other subjects attacked there were fragmentation studies of $\nu_{\mu}$p, $\nu_{\mu}$n, $\overline{\nu_{\mu}}$p and $\overline{\nu_{\mu}}$n CC interactions, analyses of specific exclusive channels, analyses of the transverse momenta distributions and also a search for fractionally charged particles which could have been produced in (anti)neutrino-deuterium interactions.

\begin{figure}[h]
\begin{center}
\hspace{-0.3cm}
{\centering\resizebox*{!}{6cm}{\includegraphics{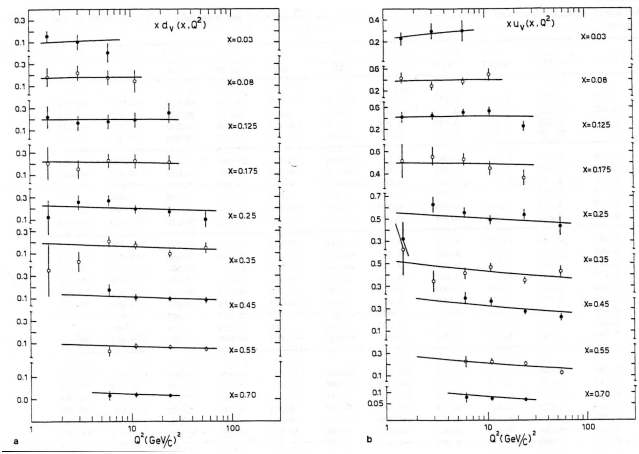}}\par}
\caption{\small a) b) $x$ and $Q^{2}$ dependence of the $xu_{v}$ and $xd_{v}$ structure functions. The curves represent the results of the QCD fits obtained with $W^{2}>$3 GeV$^{2}$, using $F^{vn}_{2}/2$ and $F^{vp}_{2}/2$ for $x>0,4$.}
\label{fig:5}
\end{center}
\end{figure}

\begin{figure}[h]
\begin{center}
{\centering\resizebox*{!}{12cm}{\includegraphics{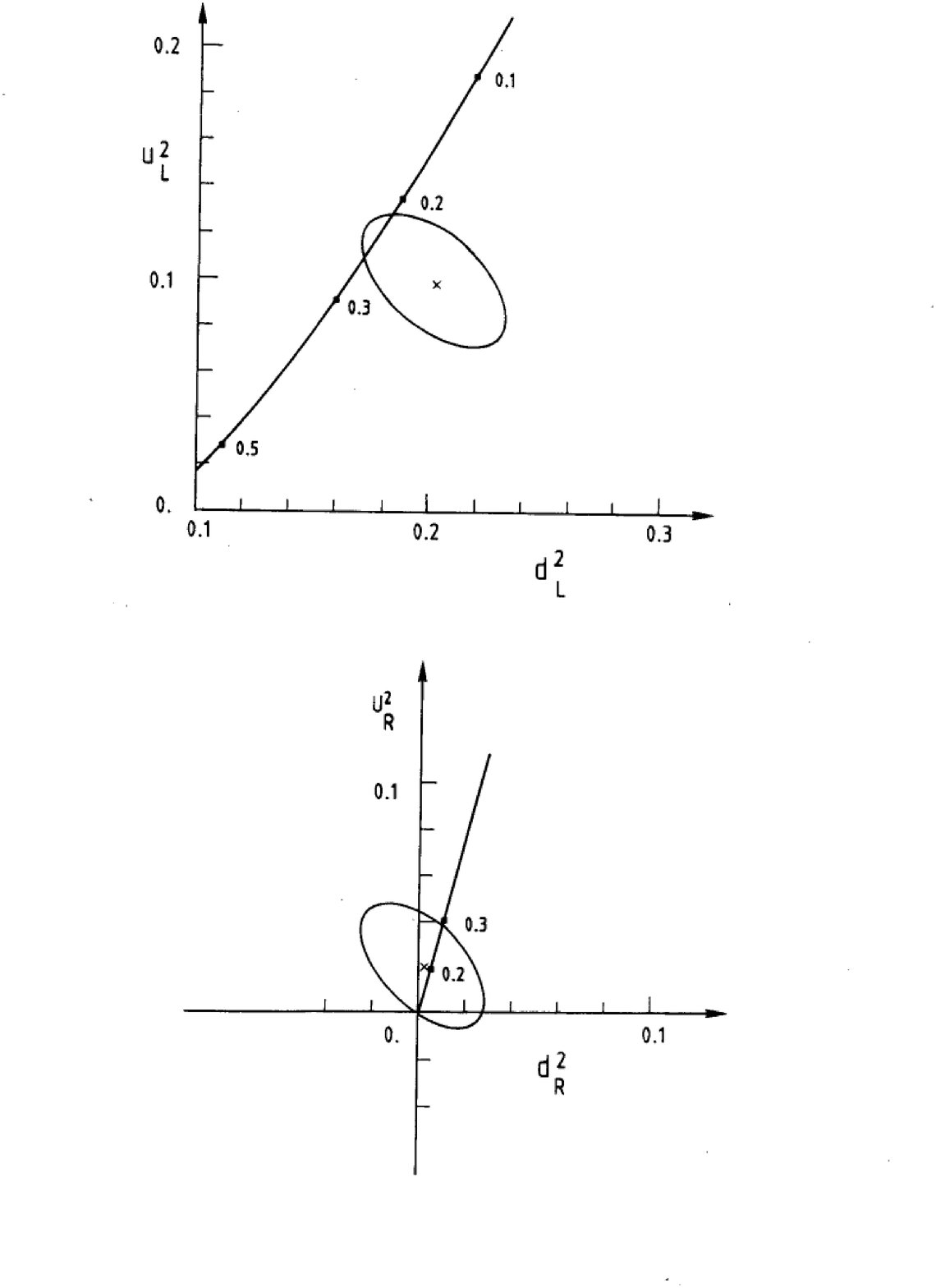}}\par}
\caption{\small Left handed, $u^{2}_{L}$ vs $d^{2}_{L}$, and right handed, $u^{2}_{R}$ vs $d^{2}_{R}$, coupling constants. The one standard deviation contours correspond to our data. The lines correspond to the electroweak model predictions as a function of $sin^{2}\theta_{W}$.}
\label{fig:6}
\end{center}
\end{figure}

\begin{figure}[h]
\begin{center}
\hspace{-1.1cm}
\vspace{-0.5cm}
{\centering\resizebox*{!}{4.1cm}{\includegraphics{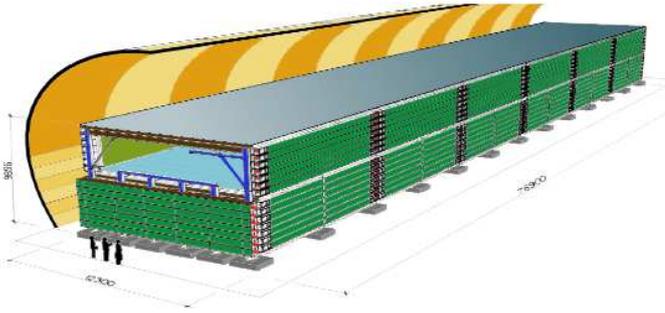}}\par}
\vspace{0.5cm}
\caption{\small Layout of the Macro experiment in hall B of the underground Gran Sasso Lab.}
\label{fig:7}
\end{center}
\end{figure}

\begin{figure}[h]
\begin{center}
\hspace{-0.5cm}
{\centering\resizebox*{!}{6.5cm}{\includegraphics{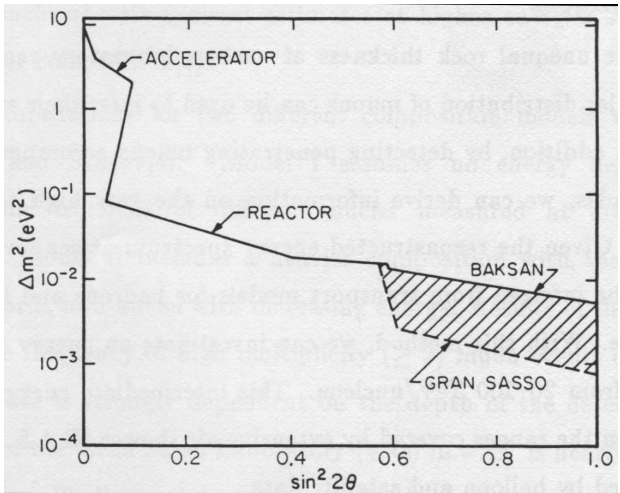}}\par}
\caption{\small Photocopy of the graph in page 31 of the 1984 MACRO proposal which summarized the situation of neutrino oscillation limits in 1984 and stated the goal for the MACRO atmospheric neutrino oscillation searches (shaded region) \cite{35}, which were actually observed there! \cite{29}.}
\label{fig:8}
\end{center}
\end{figure}

\section{The MACRO experiment at Gran Sasso.}

Fig. 6 gives a global view of the MACRO detector in the underground hall B of the Gran Sasso lab. The experiment was made of 6 supermodules, each with a lower and an upper part (the Attico, which contained the electronics). It was made with 3 horizontal layers of liquid scintillators for timing and 14 layers of limited streamer tubes for tracking; it was covered on all sides by one layer of liquid scintillators and 6 layers of limited streamer tubes to make a closed box. It also had 1 horizontal and 1 vertical layer of Nuclear Track Detectors \cite{28}. 

This multipurpose experiment was planned for atmospheric muon studies, searches for GUT monopoles, nuclearites and for antineutrinos from stellar collapses, studies of atmospheric $\nu_{\mu}$ oscillations, and others [28-35]. Flaminio participated in the proposal, in the general planning, in the construction of the first lower supermodule and the first running period. Fig. 7 shows a photocopy of the proposal concerning the status of neutrino oscillation searches and the limits existing in 1984, and a statement on where MACRO would contribute, the dashed region in Fig. 7 \cite{35}: it was exactly there that MACRO found the atmospheric neutrino oscillations. This was the major discovery of MACRO and the final parameters of the oscillations are $\Delta$m$_{23}^{2}$=2.3 $\cdot$ 10$^{-3}$ eV$^{2}$ , sin$^{2}$$2\theta$$_{23}$=1 \cite{29}, well in agreement with the results of other experiments (Soudan2 and Superkamiokande, and later the long baseline experiments KeK and Minos). The world averages are now $2.4 \cdot 10^{-3}$ eV$^{2}$ and maximal mixing, t.i. sin$^{2}$$2\theta$$_{23} \simeq 1$, respectively \cite{36}. The agreement among the different experiments is well inside their statistical and systematic uncertainties. 

The other major results of Macro concerned the points discussed at the beginning of the previous paragraph and may be found in Ref. [30-34].

\section{Conclusions. Perspectives.}

The bubble chamber experiments recalled above contributed to our present understanding of the microworld \cite{36}.

Selected bubble chamber pictures provided an intuitive view of particle physics phenomena and proved that our field is based on simple and intelligibile experimental facts \cite{27}.

The relatively new field of astroparticle physics started seeing some new phenomena, like the discovery of neutrino oscillations \cite{29}, that could open the door to Physics Beyond the Standard Model of Particle Physics.

The collaboration with Enzo Flaminio is continuing using neutrino telescopes \cite{1, 37, 38}.

It was and it is a pleasure to collaborate with Enzo and I hope that this continues for ... a long time!\\

{\bf Acknowledgements.} 
\\

I thank several colleagues for discussions and advices. I thank drs. M. Errico and M. Giorgini for corrections and technical support.


\begin{thebibliography}{4}

\bibitem{1} M. Anghinolfi, "Antares and km3", this symposium.
\bibitem{2} Bubbles 40, Nuclear Phys. B (Proc. Suppl.) 96 (1994).
\bibitem{3} G. Giacomelli, physics/0604152; http:www.bo.infn.it
\bibitem{4} A. Bigi et al., Nucl. Phys. B110 (1976) 25.
\bibitem{5} W. Cameron et al., Nucl. Phys. B132 (1978) 189.
\bibitem{6} S.I. Alekhin, et al. CERN-HERA 87-01 (1987).
\bibitem{7} D. Allasia et al., Phys. Lett. 107B (1981) 148.
\bibitem{8} S. Barlag et al., Z. Phys. C-Particles and Fields 11 (1931) 283.
\bibitem{9} D. Allasia et al., Phys. Lett. 1178 (1982) 262.
\bibitem{10} D. Allasia et al., Phys. Lett. 124B (1983) 543.
\bibitem{11} D. Allasia et al., Nucl. Phys. B224 (1983) 1.
\bibitem{12} D. Allasia et al., Phys. Lett. 133B (1983) 129.
\bibitem{13} D. Allasia et al., Phys. Lett. l35B (1984) 231.
\bibitem{14} D. Allasia et al., Z. Phys. C24 (1984) 119.
\bibitem{15} D. Allasia et al., Nucl. Phys. B239 (1984) 301.
\bibitem{16} D. Allasia et al., Zeit. Phys. C27 (1985) 239.
\bibitem{17} D. Allasia et al., Phys. Lett. 154B (1985) 231.
\bibitem{18} D. Allasia et al., Z. Phys. C28 (1985) 321.
\bibitem{19} D. Allasia et al., Phys. Rev. D31 (1985) 2996.
\bibitem{20} D. Allasia et al., Nucl. Phys. B268 (1986) 1.
\bibitem{21} D. Allasia et al., Phys. Rev. D37 (1988) 219.
\bibitem{22} D. Allasia et al., Z. Phys. C37 (1988) 52.
\bibitem{23} D. Allasia et al., Nucl. Phys. B307 (1988) 1.
\bibitem{24} B. Jongejans et al., Nuovo Cimento 101A (1989) 435.
\bibitem{25} D. Allasia et al., Nucl. Phys. B343 (1990) 285.
\bibitem{26} J. Guy et al., Z. Phys. C36 (1987) 337.
\bibitem{27} G. Giacomelli, La Fisica nella Scuola 19 (1986) 173.\\
             F. Fabbri, et al., CERN Courier july/august 2003 page19.
\bibitem{28} M. Calicchio et al. Nucl. Instrum. Meth. A264 (1988) 18.\\
             M. Ambrosio et al., Nucl. Instrum. Meth. A486 (2002) 663.
\bibitem{29} M. Ambrosio et al., Phys. Lett. B434 (1998) 451.\\
             M. Ambrosio et al., Phys. Lett. B517 (2001) 59.\\ 
             M. Ambrosio et al., Eur. Phys. J. C36 (2004) 323.
\bibitem{30} G. Battistoni et al., Phys. Lett. B615 (2005) 14.
\bibitem{31} M. Ambrosio et al., Eur. Phys. J. C25 (2002) 511.
\bibitem{32} M. Ambrosio et al., Eur. Phys. J. C37 (2004) 265.
\bibitem{33} M. Ambrosio et al., Astropart. Phys. 20 (2003) 145.\\
             S. Cecchini et al., Eur. Phys. Lett. 87 (2009) 39001.
\bibitem{34} M. Ambrosio et al., Astrophys. J. 546 (2001) 1038.
\bibitem{35} C. De Marzo et al., (The MACRO proposal) CALT-68-1237 (1984).
\bibitem{36} C. Amsler et al. (P.D.G.), Phys. Lett. B667 (2008) 1.\\ 
G. Giacomelli, "The Standard Model. Neutrino Oscillations", Rad. Meas. 44 (2009) 826, arXiv:0901.2492v2 [hep-ex].  
\bibitem{37} M. Ageron et al., Nucl. Instrum. Meth. A581 (2007) 695.\\
             P. Amram, et al., Nucl. Instrum. Meth. A484 (2002) 369.
\bibitem{38} E. Migneco, et al., Nucl. Instrum. Meth. A567 (2006) 444.\\
             S. Aiello et al., Astropart. Phys. 28 (2007) 1.

\end{thebibliography}
\end{document}